\documentclass[conference,keeplastbox]{IEEEtran}
\IEEEoverridecommandlockouts
\pdfoutput=1
\usepackage[utf8]{inputenc}
\usepackage{longtable}
\usepackage{makecell}
\usepackage{graphicx}
\usepackage{textcomp}
\usepackage[dvipsnames]{xcolor}
\usepackage{cite}
\usepackage{enumitem}
\usepackage{verbatim}
\usepackage{multirow}

\usepackage{amsmath,amssymb,amsfonts}
\usepackage[ruled,vlined]{algorithm2e}

\usepackage[normalem]{ulem}
\usepackage{subcaption}

\SetCommentSty{mycommfont}

\usepackage{tikz}
\usepackage{pgfplots, pgfplotstable}

\usepackage[textsize=tiny,textwidth=1.4cm]{todonotes}

\usetikzlibrary{scopes}
\usetikzlibrary{calc,chains,positioning,shapes.geometric,shadows}

\usepackage{url}
\usepackage[capitalise,noabbrev,nameinlink]{cleveref}
\usepackage{listings}

\begin{document}
\setlength{\marginparwidth}{1.38cm}

\title{\textsc{Transit-Gym}: A Simulation and Evaluation Engine for Analysis of Bus Transit Systems
\thanks{* Both Rongze Gui and Ruixiao Sun contributed to the paper equally.}}

\makeatletter
\newcommand{\linebreakand}{%
  \end{@IEEEauthorhalign}
  \hfill\mbox{}\par
  \mbox{}\hfill\begin{@IEEEauthorhalign}
}
\makeatother


\author{\IEEEauthorblockN{Ruixiao Sun\IEEEauthorrefmark{1}, Rongze Gui\IEEEauthorrefmark{2}, Himanshu Neema\IEEEauthorrefmark{2}, Yuche Chen\IEEEauthorrefmark{1}, Juliette Ugirumurera\IEEEauthorrefmark{3}, \\
 Joseph Severino\IEEEauthorrefmark{3}, Philip Pugliese\IEEEauthorrefmark{4},
Aron Laszka\IEEEauthorrefmark{5}, Abhishek Dubey\IEEEauthorrefmark{2}
}
\IEEEauthorblockA{\IEEEauthorrefmark{1}
University of South Carolina \IEEEauthorrefmark{2}
Vanderbilt University \IEEEauthorrefmark{3}
National Renewable Energy Lab
}
\IEEEauthorblockA{\IEEEauthorrefmark{4}
Chattanooga Area Regional Transportation Authority
 \IEEEauthorrefmark{5}
University of Houston
}
}

\maketitle

\begin{abstract}
Public-transit systems face a number of operational challenges: (a) changing ridership patterns requiring optimization of fixed line services, (b) optimizing vehicle-to-trip assignments to reduce maintenance and operation codes, and (c) ensuring equitable and fair coverage to areas with low ridership. Optimizing these objectives presents a hard computational problem due to the size and complexity of the decision space. State-of-the-art methods formulate these problems as variants of the vehicle routing problem and use data-driven heuristics for optimizing the procedures. However, the evaluation and training of these algorithms require large datasets that provide realistic coverage of various operational uncertainties. This paper presents a dynamic simulation platform, called \textsc{Transit-Gym}, that can bridge this gap by providing the ability to simulate scenarios, focusing on variation of demand models, variations of route networks, and variations of vehicle-to-trip assignments. The central contribution of this work is a  domain-specific language and associated experimentation tool-chain and infrastructure to enable subject-matter experts to intuitively specify, simulate, and analyze large-scale transit scenarios and their parametric variations. Of particular significance is an integrated microscopic energy consumption model that also helps to analyze the energy cost of various transit decisions made by the transportation agency of a city. 
\end{abstract}

\section{Introduction}
\label{sec:introduction}

Public-transit systems face a trade-off between concentrating service into high-utilization routes that serve large numbers of people and spreading out service to ensure that people everywhere have access to at least some service. As a result, improving the efficiency of an existing system while enhancing service in terms of usefulness and coverage is challenging. The problem of transit systems is often treated as an integrated dynamic optimization problem focusing on three objectives: minimizing energy per passenger per mile, minimizing total energy consumed, and maximizing the percentage of daily trips served by public transit\cite{phillips2004application,murray2003coverage,lai2011behavioral}. The last objective often requires dividing the transit area into fixed line and on-demand routes. 

Optimizing integrated transit services presents a very challenging computational problem. Firstly, the integrated operational decision space is vast: it includes dispatch and routing of on-demand service vehicles, scheduling and placement of flexible courtesy stops for fixed-route service, and deciding the allocation of a mixed transit fleet to specific routes, i.e., deciding which route should be assigned an electric vehicle and which route should be assigned to diesel vehicles\cite{sayarshad2020optimizing,sivagnanam2021minimizing}. Further, for fleet with electrical vehicles, it is also necessary to strategically schedule the charging slots to ensure that there is no undue burden on the electric grid. 
While it is possible to optimize these decisions separately as prior work has done, integrated optimization can lead to significantly better service (e.g., synchronizing flexible courtesy stops with on-demand transit (microtransit) dispatch for easy transfer). Further, decisions must be made facing uncertainty (e.g., future demand requests and traffic conditions, unscheduled maintenance). Despite these uncertainties, transit services must meet strict requirements: paratransit requests must be served within a limited time frame, fixed-route schedules must be closely followed, etc. Finally, the operational environment is evolving due to changes in population, traffic habits, etc.

Our team has been developing state-of-the-art artificial intelligence, machine learning, and data-driven optimization techniques for solving these challenges. This includes development of microscopic and macroscopic energy estimation models for assessing the cost of running the transit vehicles \cite{gallet2018estimation,ayman_data}. However, till now, our and similar efforts from other research teams have suffered from a bottleneck. They can only test their models against previously collected data, which often does not cover all possible operational scenarios. Our hypothesis is that this challenge can be addressed by using  simulation engines that use publicly-available street maps, transit network information, and past transit performance and enable realistic simulation for varying scenarios. Such simulations serve two purposes: (1) enable generation of new datasets that augment real-world data and (2) enable evaluation of algorithms in terms of transit demand met and overall energy cost.

In this paper, we describe \textsc{Transit-Gym}, a SUMO based general-purpose transit simulator, which we demonstrate by using real-world scenarios and calibrated data from Chattanooga, TN. This simulator extends a transportation simulator developed by the team, called Mobilytics-Gym (built using MATSIM) \cite{Samal2019}, which integrates agent behavior models that can be learned from surveys and real-word data with transit infrastructure models (street maps and static transit data).  The challenge in making the simulation realistic is ensuring that the simulated environment reflects the real-world travel time delays and weather scenarios.
This has been achieved by careful calibration of the underlying model. Note that in order for the simulator to remain viable, it is crucial to keep the physical transit network of the city and the simulated transit network in sync. The geometric design of roads evolves over time as new roads and additional lanes are built on existing transportation infrastructure. This adds a layer of complexity to developing precise simulation environments. It is common for developers and researchers to procure transportation data and network files from existing platforms, such as OpenStreetMap (OSM)\cite{haklay2010good}. The data available on these platforms is traditionally crowdsourced, and in many cases, it is not up to date with the actual transportation infrastructure.
Therefore, we have developed and integrated specialized procedures to perform a stochastic check against the designed network and real traffic data collected from the city and then fixing any discrepancies in the network. This process ensures that that road geometries and map files are continuously updated, providing an accurate environment for testing and validation.

The key contributions of this paper are as follows:
\begin{itemize}
    \item A novel DSML that allows intuitive specification and variation of transit scenarios.
    \item A methodology to construct and calibrate street maps that conform to real-world transportation infrastructure.
    \item A method to calibrate travel demand both for the road traffic as well as on transit services.
    \item A toolchain that automatically configures SUMO simulations from the scenarios specified using the above DSML.
    \item A customized general-purpose SUMO simulation specifically for transit scenarios.
    \item A novel method to perform detailed energy consumption of transit buses.
    \item An analytical design to produce experiment data that is amenable for data analytics.
    \item An integrated dashboard to visualize the experiment results in an highly intuitive manner and one that focuses on key operational metrics.
    \item An integrated highly scalable cloud backend for transit simulations.
\end{itemize}

The rest of the paper is organized as follows: Section~\ref{sec:background} provides a background on key concepts used in construction of \textsc{Transit-Gym}. Section~\ref{sec:methodology} provides a detailed overview of our approach to transit simulation. Section~\ref{sec:experimentResults} presents scenarios and experiment results that demonstrate the system's usability. Finally, Section~\ref{sec:conclusions} concludes the paper and presents directions for future work.

\section{Background}
\label{sec:background}

\subsection{Model Integrated Computing}
Model-Integrated Computing (MIC) \cite{sztipanovits1997model} is a meta-modeling technique that puts models at the center of all applications. MIC allows the creation of domain-specific modeling languages (DSML) called metamodels. Using the DSML, domain-specific models are created based on it representing different applications or scenarios. The DSML describes syntax and semantics of the models, how models can be constructed, and the constraints that the valid models must satisfy. The DSML tools then interpret models to verify them, generate artifacts associated with the models, and generate executables that are executed per the DSML's semantics.

Aspect-oriented programming (AOP) \cite{kiczales1996aspect} is related technique where integrated systems are composed by weaving different systems written in imperative languages, whereas MIC focuses on model-based application synthesis using DSMLs. Existing GenVoca \cite{batory1997composition} generators compose systems via generation of systems in abstraction layers using object-oriented techniques, but do not support multiple-aspect composition. Perhaps a well-known related effort is Model-Driven Architecture (MDA) \cite{mda} that uses UML for modeling, whereas MIC is flexible and utilizes domain-specific modeling techniques.

MIC facilitates rapid synthesis of domain-specific applications and scenarios and enables generation of syntactic- and semantic-conformant executable artifacts. We use metamodeling by developing a DSML that is specific to transit simulations. We describe the scenario modeling language and example scenario models later in the paper.

\subsection{Cyber-Physical Systems Wind Tunnel}
Vanderbilt has developed a model-based heterogeneous simulation integration and experimentation framework known as the Cyber-Physical Systems Wind Tunnel (CPSWT)\cite{neema2019simulation}. CPSWT framework has been successfully used in different application domains such as road traffic security and resilience\cite{koutsoukos2017sure}, efficient partitioning and co-simulating high-fidelity dynamical simulations\cite{neema2014model}, and smart grids\cite{neema2016c2wt}. The framework is currently being actively developed at Vanderbilt and was recently adopted by NIST, where it is being further enhanced for practical applications in collaboration with Vanderbilt\cite{burns2018universal}. We are working on extending \textsc{Transit-Gym} using the CPSWT framework for integrating traffic simulation in SUMO with transportation planning simulation and for using CPSWT courses-of-action (COA) evaluation method\cite{neema2018integrated} for large-scale scenario-based experimentation.

\subsection{SUMO}
\textsc{Transit-Gym} uses SUMO as its main road traffic and transit simulator. SUMO\cite{krajzewicz2012recent} is a highly customizable, microscopic traffic simulator. It uses a discrete-time computation model and is highly scalable to city-level simulations. In SUMO, arbitrary maps can be designed as well as real-world maps can be imported from OSM\cite{haklay2010good}. We chose SUMO in \textsc{Transit-Gym} because it provides highly detailed capability for modeling transit demand and routes, it is much easier to configure the high-fidelity simulations, and provides detailed traffic and transit experiment results.

In literature, SUMO or similar traffic simulation software are used to analyze operation of transit services. Particularly, there are several recent literature focusing on operation analysis of shared or on-demand mobility service using SUMO or similar traffic simulation software\cite{aziz2020optimization,becker2020assessing,zhu2021shared}. However, these studies’ traffic simulation is on limited geographic area and limited fleet size, which is normally the case for shared mobility services. Very limited studies focus on simulation of transit fleet at a whole urban area. Particularly, to the best of authors’ knowledge, there is no prior study that utilizes model integrated computing framework to conduct scenario-based transit simulations on a city scale with real-world background traffic data. This paper tries to fill in this knowledge gap.

\subsection{Energy Estimation Models}
Existing energy prediction models for electric vehicles can be classified as microscopic and macroscopic models according to prediction resolution. Macroscopic models utilize aggregated traffic and vehicle information, over a time period, e.g. 1 hour, to estimate energy consumption of electric vehicle \cite{gallet2018estimation,masikos2015mesoscopic,pamula2020estimation,sun2019machine}. Although the macroscopic models are interesting and have solid methodologies, they can only provide energy estimation over a time period, thus they are not useful in evaluating energy savings of real-time bus operation strategies,
which have time-resolution only of a few seconds.

Microscopic models can estimate energy consumption of electric vehicle at high frequency levels, and are widely adopted in real-time optimal control of traffic involving electric cars\cite{genikomsakis2017computationally,luin2019microsimulation}.
Very limited literature studied electric bus. Relevant electric bus studies adopted physical-based approaches to estimate electricity consumption. But these physical based microscopic models lack transfer ability and cannot be easily used in other vehicles. In a previous study, we have developed machine-learning based microscopic energy estimation model that takes 1Hz vehicle trajectory data as input and output 1Hz energy consumption estimation\cite{sun2021hybrid}.

\section{Methodology \& Scenario Construction}
\label{sec:methodology}

\subsection{Simulation Platform Setup \& Data Sources}
Figure~\ref{fig:framework} presents the framework of scenario-based simulation platform, which leverages the open-source simulator SUMO\cite{krajzewicz2012recent}, DSML and energy estimation model. SUMO is used to conduct microscopic traffic simulation, wherein vehicles, public transport and persons are modelled explicitly. The DSML is developed to provide an interpreter for customizing transit simulation configurations based on scenario specifications. Energy estimation models are constructed on a machine learning algorithm, which is used as a tool to predict the energy consumption for each trajectory simulated bus. We now discuss the framework in detail.

\begin{figure}[ht]
	\centering
	\includegraphics[width=.45\textwidth]{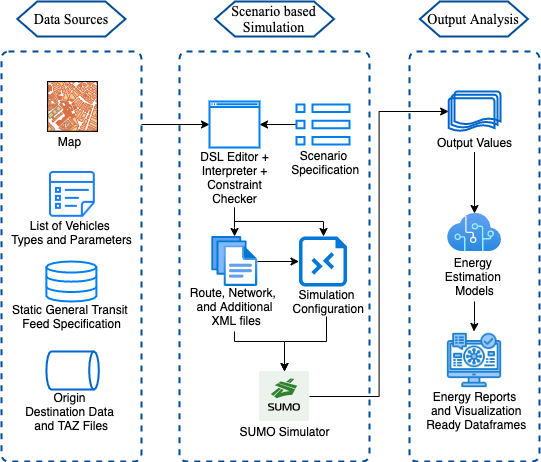}
	\caption{Framework of Simulation Platform}
	\label{fig:framework}
\end{figure}

The data sources used in this simulation include:
\begin{itemize}

    \item \textbf{Map}: The map data can be generated from Open Street Maps (OSM)~\cite{haklay2010good}. OSM is a collaborative project that provides fine calibrated map data. With the OSM map data, we used the SUMO built-in package NETCONVERT to convert the OSM map to SUMO loadable road networks. The road network contains information such as the number of lanes in a street, speed restrictions and the traffic direction.
    
    \item \textbf{List of Vehicle Types and Parameters}: The specifications of vehicles that are employed in the city public transit system are listed as an input data for SUMO simulation. Based on such list, a vehicle type definition file will be created, containing the parameters such as max speed, acceleration, passenger capacity, min gap, driver preferences, etc.
    
    \item \textbf{Static General Transit Feed Specification (GTFS)}: The GTFS data contains the transit route information and the bus stop locations along the route, as well as the trip schedules for a day. The DTFS data will be utilized to configure the bus stops definition and bus trip schedules for transit simulation.
    
    \item \textbf{Origin Destination (OD) Data and TAZ Files}: The OD matrix data used in the platform contains details of trips through traffic analysis zone (TAZ). Each  OD  matrix  cell  is associate  with  a  daily  time  period  and  describes  the respective  traffic  demand  from  an  origin  area  zone  to  a  destination  area  zone. The TAZ files contain shape files that define region TAZ, which can be converted to SUMO TAZ with edges assigned to TAZ via POLYCONVERT tool in SUMO. With these OD matrix data and TAZ definition, we can generate the SUMO loadable person plans and daily trips for other vehicles, e.g. passenger cars and trucks.

\end{itemize}

In the following paragraphs, we discuss the details of how bus mode is implemented in the simulation. Figure~\ref{fig:simulation-workflow} presents a workflow diagram for the public transit simulation, demonstrating the interactions between the various components of SUMO. Below are the key steps in the workflow:
\begin{enumerate}

    \item \textbf{Network Refinement}: Due to the frequent mismatch between available input data and the necessary level of detail for microscopic simulation, network converted from OSM directly commonly missing or mismatching some road components. Therefore, after converting the network, network refinement is needed. NETEDIT is a graphical network editor which can be used to create, analyze and edit network files. This serves to complement the network generation heuristics with manual refinements.
    
    \item \textbf{Bus Stop Location Retrieving}: GTFS provides the geographic coordinates of bus stops. Based on such GPS coordinates, we can retrieve the location information of stops (i.e., which point of the edge ID alone with lane ID the stop is located on) from the network via TraCI~\cite{wegener2008traci}. TraCI is a SUMO module, which is capable of online interaction, for example, it allows a python script to extract various values from the simulation or change certain simulation parameters which cannot be done using the basic SUMO files. 
    
    \item \textbf{Bus Trip Generation}: Using GTFS and bus stop location as the input one can define a schedule for vehicles on their routes and so provide a scheduled public transport. The interpreter serves the function to automatically generate XML file for bus trip definition based on sequential bus stops along each bus line in the GTFS data.

    \item \textbf{Vehicle Type Definition}: The interpreter provides the function to automatically generate codes for vehicle type definition file according to the list of vehicle types and associated parameters.
    
    \item \textbf{Background Daily Traffic Route Generation}: The background traffic to the transit operation was generated from a 2014 daily average OD matrix that was provided by the Chattanooga-Hamilton County regional planning agency. This OD matrix describes the demand in a table, which the number of passenger vehicles, single unit trucks, and trailer trucks per hour originating from an origin TAZ to a destination TAZ. The OD matrix was then converted into passenger vehicle, truck, and trailer trips using the SUMO module OD2TRIPS. The resulting traffic forms the background traffic to the transit traffic. 
    
    \item \textbf{Person Trip Generation}: As mentioned above, the SUMO OD2TRIPS tool provides a way to convert OD matrix data along with TAZ into individual trips. In order to generate the departure time for each vehicle/person, either uniform or random distribution within a given time period can be chosen. By default, OD2TRIPS generates vehicular traffic, such as passenger cars and trucks. The person trip with public transit mode can also be generated by adding an extra option (--persontrips) during the OD2TRIPS executing. 
        
    \item \textbf{Bus and Person Integrated Route Generation}: The design of the transit simulation is focus on the route for bus with person riding, with the routes for other traffic modes served as the background traffic. For bus mode traffic, SUMO’s DUAROUTER software offers already intermodal routing which could be used to generate the bus route with person plans on transit. This way it suffices for the user to incorporate person plans with start and arrival points as well as the available bus schedules and the depart time and SUMO will find the fastest routes for person travel via public transport. Together with vehicle to trip assignment scenario this can give a very pseudo-realistic scenario with walking and riding persons. This process will be called in DSML Interpreter automatically.
    
    \item \textbf{Additional Settings}: To get the road segment level traffic measurement result, an edge-based state dump is defined within an additional-file served as an additional querying in the sumo configuration. In addition, the measurement values can be aggregated by setting the aggregation period. The Interpreter offers this parameter flexible augment.

\end{enumerate}

As shown in Figure~\ref{fig:simulation-workflow}, the DSML interpreter automate the processes including calling SUMO packages and writing program to generating sumo essential input files, as well as other additional settings and augments are functioned in. 

\begin{figure}[ht]
	\centering
	\includegraphics[width=\columnwidth]{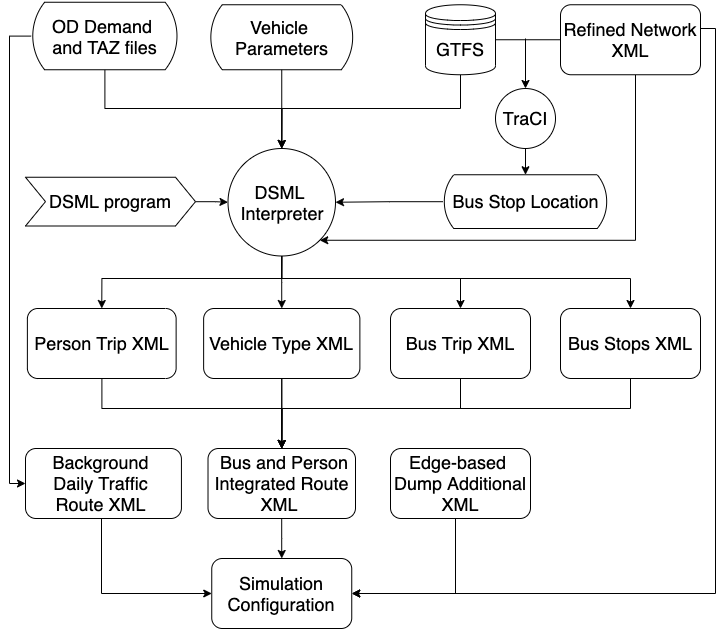}
	\caption{Workflow of the Public Transit Simulation}
	\label{fig:simulation-workflow}
	\vspace{-.1in}
\end{figure}

\subsection{Scenario-Based Simulation}
To evaluate different policies, our goal lies in simulating the exact environment they represent. Moreover, to generate sufficient datasets for algorithm training, this means of simulation should be dynamic enough to adjust to different scenarios in an efficient and timely manner. \par Our solution to this problem is the developed scenario-based simulation, where in this framework one or more policies can be quickly converted to program-readable scenarios. Those scenarios, represented as DSML code snippets, would then be sent to the interpreter to be processed.
For the detailed progress of scenario-based simulation, three components are supported:

\begin{itemize}

    \item \textbf{Scenario Specification}: For a given scenario, the road network, the list of vehicles and types and associated vehicle parameters, as well as OD matrix data along with TAZ SUMO file will be imported into the DSML interpreter. In addition, the latest GTFS data and bus stops definition will be queried directly in order to configure the bus schedules and bus stops for transit services. The detailed procedures calling SUMO tools to set up public transit schedule within the DSML interpreter scripts are shown in Figure~\ref{fig:simulation-workflow}. Furthermore, the block and vehicle to trip mapping constraints ensure that the vehicle assignments conform to transit block rules. The constraints are set to make sure that user created scenario models are valid. For example, bus assignments should be consistent across a given block. The capability to generate SUMO loadable files enables rapid changes to scenarios.
    
    \item \textbf{Configuration Generation}: The interpreter takes all of the input information along with the scenario model to generate the needed SUMO files for the simulation, which includes the road network, route files, and bus stop definition file, as well as other additional sumo files. Next, the interpreter writes the specified SUMO configuration .sumo.cfg  file, containing all the generated input files, simulation parameters such as start and end time, declaimed output values, and other settings such as waring and error log report.
    
    \item \textbf{Simulation Run}: The simulation can be executed directly with the single configuration .sumo.cfg file via command line. The interpreter also can support the execution of the simulation if chosen by the user to do so. The framework also plans to run the simulation in parallel for multiple scenarios on a cloud-computing platform, in addition to speed up the running time for large-scale simulation-based experiments.
    \item \textbf{Energy Estimation}: The output of traffic simulation include driving trajectories of buses. The trajectories contain vehicle instantaneous speed and acceleration at 1Hz frequency for each bus trip. We have developed artificial neural network (ANN) based microscopic energy estimation model that takes 1Hz vehicle trajectory data as input and output 1Hz energy consumption estimation. The microscopic energy estimation model was trained and tested on the same bus fleet operated by Chattanooga Area Regional Transportation Authority. Detailed information of the model can be found in literature \cite{sun2021hybrid}. After energy estimation for each trip, we aggregate 1Hz energy consumption into the whole trip and divide trip distance by total energy consumption to obtained energy economy for diesel and electric bus trips. For comparing, we converted electricity consumption to diesel equivalent gallon using 3600 kJ for 1 kWh and 146,520 kJ for 1 diesel equivalent gallon.

\end{itemize}

\subsection{Scenario Construction}
As mentioned before, the scenario construction is performed through the use of DSML developed on the textual metamodeling framework TextX and used in simulation configuration. The  DSML  interpreter  provides an automatic way for parameter flexibility on vehicle type, scenario generation, and block and vehicle to trip mapping constrains checker in customizable SUMO simulation.

\par A DSML program is separated into the import section and configure section. The first section imports the pre-configured scenario settings that would be used as background data for subsequent code. As explained in the last section, the pre-configured data source include Map, Vehicle Type, GTFS and OD matrix Data and TAZ Files. The import command has four variations corresponding to each of the imports. \textit{import network} configures the map, \textit{import vehicle} configures the vehicle type, \textit{import gtfs} configures the GTFS data and \textit{import td} configures transportation demand files. The next section configures one or more simulation based on user-input parameters. Here, first the time range for the simulation will be specified. The schedule command, then, is used to configure the schedule to be simulated (weekend or weekday). The output\_sampling\_period specifies the period for edge summary output. Finally, in the vehicleassignment block, one or more assignments can be defined, which associate vehicle type to blocks and/or trips based on their number. When executed, this program generates the outputs to be processed subsequently by the energy estimation model. For example, we can simulate a scenario with one weekday traffic and one weekend traffic in Chattanooga before 12am, and assign a specific vehicle type to all bus routes within block 101. Listing~\ref{lst:dsmlProgram} shows part of an example program that constructs this scenario.

\begin{center}
\begin{lstlisting}[linewidth=1\linewidth, basicstyle=\small\ttfamily\bfseries, stringstyle=\color{green!50!black}, keywordstyle=\bfseries\color{blue}, language=Python, morekeywords={simulation}, breaklines=true, frame = trBL, caption={Example DSML program}, captionpos=b, label=lst:dsmlProgram]
import "network.Chattanooga"
import "vehicle.BUS_type.xlsx"
import "gtfs.latest"
import "td.OD_person.od"

simulation configuration 1 {
    time [0000:1200]
    schedule weekday
    output_sampling_period 3600
    vehicleassignment {
        block 101: "Gillig_103"
    }
}

simulation configuration 2 {
    ...
\end{lstlisting}
\end{center}

\section{Results and Discussion}
\label{sec:experimentResults}


\subsection{Simulation Output Analysis}
The platform is capable of generating various visualization ready dataframes converted from output XML files, such as edge based summary, vehicle trajectory, and bus stop information. Edge-based traffic measurement output, values within this output describe the macroscopic values such as the mean speed, the mean density, the mean occupancy of edge during specified time interval. Bus stop output contains the information about each bus simulated schedule: time of arrival and departure, stopping place and number of persons that were loaded and unloaded at each stop. Trajectory output includes information about type, current speed and acceleration of each vehicle. These output values along with vehicle to trip assignments can be used for energy analysis. In addition, the analysis results such as vehicle occupancy, boarding, alighting, OSM segment-level speed of vehicles, and segment-level travel times and congestion observed besides the buses’ trajectories are stored into data frames. It provides flexibility for users to select point of interest from these data frames and conduct specific analysis. Some examples of the output analysis visualizations are shown as following .

Figure~\ref{fig:occupancy by trip} shows the passenger occupancy of each bus along its stops by trips for a whole day. Each point in it indicates the amount of passenger on one bus before it arrived at a bus stop, and each box presents the maximum, 1st quartile, median, 3rd quartile and the minimum passenger amount for one trip.

\begin{figure*}[t]
	\centering
	\includegraphics[width=0.9\textwidth]{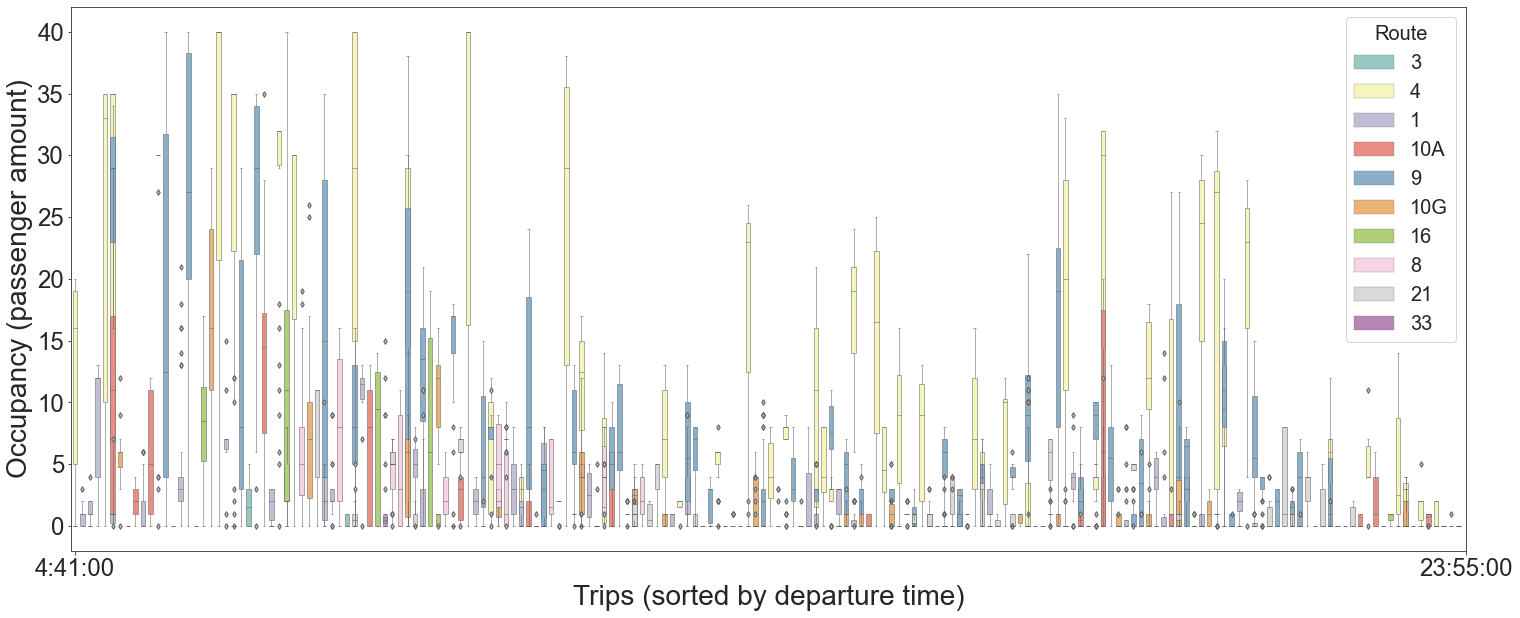}
	\caption{Passenger occupancy of each bus along the bus stops by trips across 24 hours}
	\label{fig:occupancy by trip}
	\vspace{-.1in}
\end{figure*}

Figure~\ref{fig:occupancy by route} shows the maximum passenger occupancy of each bus along the bus stops by routes. The bar within each box represents the median of the maximum passenger occupancy of different buses on each route and the two sides of box correspond to 1st and 3rd quartile of the data for each route. To investigate the occupancy status during different time in a day, the distributions of bus occupancy between three specific hours (08, 12, and 17, according to morning, midday, and afternoon) on route 4 are shown in figure~\ref{fig:occupancy dist}. 

\begin{figure}
\begin{subfigure}[ht]{0.45\linewidth}
\includegraphics[width=\linewidth]{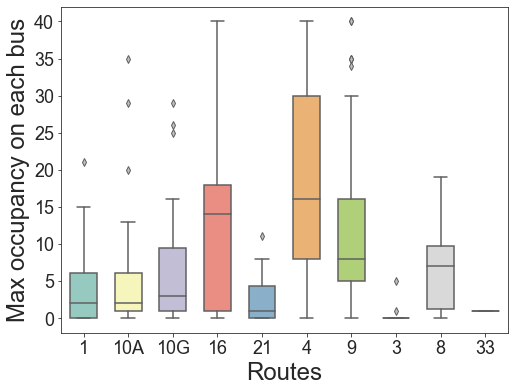}
\caption{}
\label{fig:occupancy by route}
\end{subfigure}
\hfill
\begin{subfigure}[ht]{0.5\linewidth}
\includegraphics[width=\linewidth]{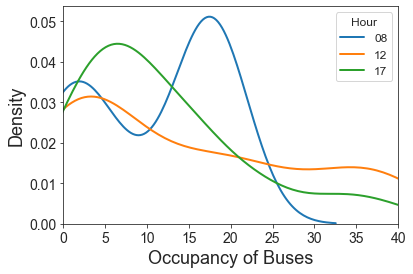}
\caption{}
\label{fig:occupancy dist}
\end{subfigure}%
\caption{Maximum passenger occupancy of each bus along the bus stops by routes across 24 hours (a). Distributions of bus occupancy between specific hours on route 4 (b)}
\end{figure}

\begin{figure}
\begin{subfigure}[ht]{0.475\linewidth}
\includegraphics[width=\linewidth]{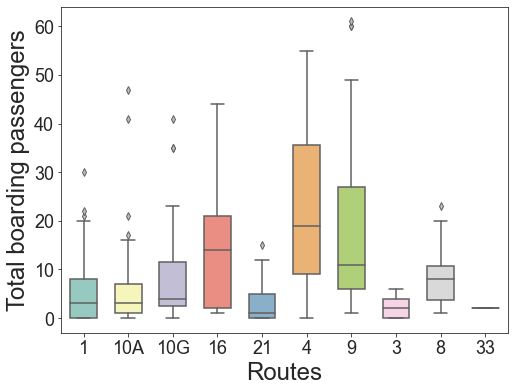}
\caption{}
\label{fig:boarding by route}
\end{subfigure}
\hfill
\begin{subfigure}[ht]{0.475\linewidth}
\includegraphics[width=\linewidth]{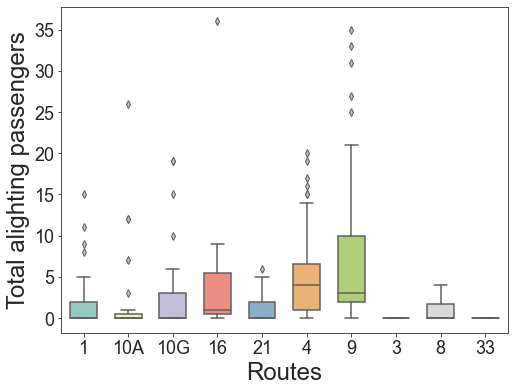}
\caption{}
\label{fig:alighting by route}
\end{subfigure}%
\caption{The total boarding passengers (a) and the total alighting passengers of each bus by routes across 24 hours (b)}
\end{figure}

Figure 5 summarizes the boarding and alighting passengers of each bus across a whole day by routes. Figure 6 shows the distributions of bus speed and average speed of buses on route 4 for three hours (08, 12, and 17 hour). The left plot implies that the bus speed during the three hours have the similar distribution and the common speeds for these three hours all lay on 20 mph. The right plot shows that the average speed of buses in morning peak hour is higher than that of midday and followed by afternoon rush hour. This reveals that the traffic is jammed on afternoon rush hour and midday, getting better at morning peak hour.

Trajectory output data for half a day was generated from our platform as an illustration for energy estimation. Figure~\ref{fig:base_scenario} shows the estimated consumption rates for buses on different routes. The range of the energy consumption rate is from 0 to 3 mile/gallon. The larger the consumption rate, the less gallon the bus consumed in the same driving distance. We can see that many buses are driving on the route 4, and some of them have higher rates than others. This means route 4 may have a better road conditions than other routes assuming all the buses are in the similar performance.

\begin{figure}
\begin{subfigure}[ht]{0.475\linewidth}
\includegraphics[width=\linewidth]{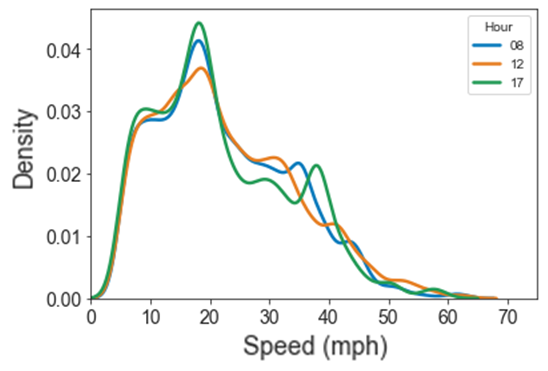}
\caption{}
\label{fig:speed dist}
\end{subfigure}
\hfill
\begin{subfigure}[ht]{0.475\linewidth}
\includegraphics[width=\linewidth]{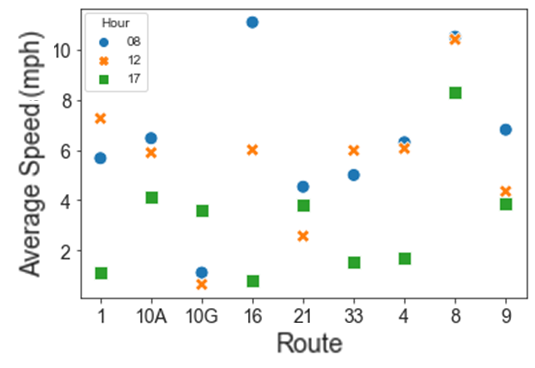}
\caption{}
\label{fig:average speed}
\end{subfigure}%
\caption{Distributions of bus speed (a), average speed of buses on route 4 during three specific hours (b)}
\end{figure}

\begin{figure}[ht]
	\centering
	\includegraphics[width=.8\linewidth]{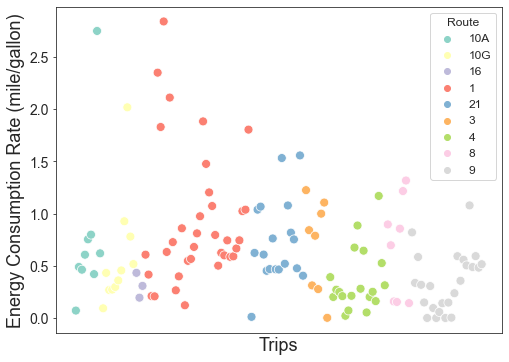}
	\caption{Energy consumption rate for buses in half a day}
	\label{fig:base_scenario}
\end{figure}

\subsection{Scenario Analysis of Energy Consumption}
This section investigates the analysis of energy consumption in different simulation scenarios, such as variations of vehicle trip assignments and variations of demand models. 

We have three types of buses, including diesel, hybrid and electric buses, as the candidates for trip assignment. The energy estimation models, which was built for training diesel, hybrid, and electric buses, are used to estimate the fuel/power used for diesel, hybrid, and electric buses accordingly.  Figure~\ref{fig:trip_assignment} presents the comparison of energy consumption rate of buses among different trip assignment scenarios, including base scenario, using all hybrid buses and using all electric buses. It is obvious that buses in the scenario using all electric trip assignment generally consumes the least energy than that of using the other assignments across all the routes.

\begin{figure}[ht]
	\centering
	\includegraphics[width=.8\linewidth]{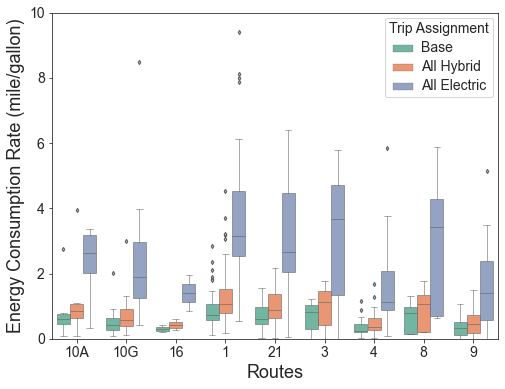}
	\caption{Energy consumption rates for buses in different trip assignment scenarios}
	\label{fig:trip_assignment}
\end{figure}

As illustrated before, the developed simulation platform has the ability to change the demand models.
To check the impact of background traffic on the energy estimation, two scenarios regarding different background demand models are constructed. The first scenario presents the simulation with the original daily background traffic, and the second scenario is the simulation with the daily background traffic reduced by 20\%. These two scenarios are conducted with in one peak hour simulation as a sample illustration. In addition, to check the ability of variation of GTFS data in our platform, the GTFS data in this comparison is on the different day against the GTFS data in the previous simulation. The estimated energy consumption rates for three bus types under the two scenarios are shown in Table~\ref{T:energy estimation}. From this table, we can see that the energy consumption for diesel buses generally greater than that of hybrid buses. The electric buses are capable of saving energy in the similar situation compared with diesel and hybrid buses. The results of original demand are higher than that of reduced demand scenario for diesel and hybrid buses, while lower than that of original demand scenario for electric buses. As shown in Figure 7, the network is in a heavy congestion at morning peak. The network with lower background trip demand has fewer jams, that is to say, the buses will have a higher speed and reduce the brake times under reduced demand scenario than the original demand scenario. For diesel and hybrid buses, fewer brakes means the bus can save fuel consumption compared with more brakes in driving. For electric bus, it has a chance to regenerate few power during braking. Therefore, it might be the reason that the electric buses consume less power under reduced demand scenario than that of original demand scenario. 

\begin{table}[!h]
\begin{center}
\caption{Estimated energy consumption rate for buses within the same period} 
\label{T:energy estimation}
\begin{tabular}{|p{0.6cm} |p{0.4cm} |p{0.6cm}|p{0.8cm}|p{0.8cm}|p{0.6cm}|p{0.8cm}|p{0.8cm}|}
\hline
\multirow{3}{2em}{\textbf{Route No.}} & \multirow{3}{2em}{\textbf{Trip No.}} & \multicolumn{6}{c|}{\bfseries\makecell[c]{Energy consumption rate\\ (mile/diesel equivalent gallon)}} \\ \cline{3-8} 
& & \multicolumn{3}{c|}{\textbf{Original demand}} & \multicolumn{3}{c|}{\textbf{Reduced demand}} \\ \cline{3-8} 
&  & \textbf{Diesel}  & \textbf{Hybrid}   & \textbf{Electric}   & \textbf{Diesel}   & \textbf{Hybrid}   & \textbf{Electric}  \\ \hline
1	& 1	& 1.0 & 1.3 & 11.6  &1.2 	&1.8 	&5.0 
\\
	& 2	& 2.7 & 4.3 & 11.7  &2.8  &4.7 	&10.7 
\\
	& 3	& 2.3 & 3.3 & 10.8  &2.7 &4.6 &10.4 
\\
	& 4	& 1.1 & 1.5 & 8.1  &1.5  &2.7 &6.3 
\\\hline
3	& 5	& 1.6 & 2.0	& 14.9  &1.9 &2.9 &9.5 
\\
	& 6	& 3.7 & 5.0	& 15.9  &4.2  &6.4 	&16.8 
\\
	& 7	& 1.2 & 1.5	& 11.4  &3.3  &5.5 	&14.6 
\\\hline
4	& 8	& 2.0 & 2.8	& 11.4  &1.4	&2.0	&5.0 
\\
	& 9	& 2.3 & 3.1	& 14.0   &2.4 	&3.7 	&10.2 
\\\hline
9	& 10 & 2.8 	& 3.8 & 14.0  &2.5 	&3.9 &10.9 
\\
	& 11 & 2.1  & 3.4 & 9.4  &2.3	&3.7  &8.0 
\\\hline
10A & 12 & 2.5	& 3.4	& 14.2  &2.5 &4.0 &10.7 
\\
	& 13 & 2.8 	& 4.2	& 12.8   &3.0 	&4.7	&10.7 
\\\hline
10G	& 14 & 2.4 	& 4.0 & 9.5  &2.8 &4.5 	&9.6 
\\
	& 15 & 0.9 	& 1.1 	& 8.1   &2.3 &3.7  &8.4 
\\\hline
\end{tabular}
\end{center}
\end{table}
\section{Conclusions \& Future Work}
\label{sec:conclusions}

Public transit services are highly complex and faces a range of operational challenges to optimize fixed-line and on-demand services in a way that minimizes cost for the riders and transportation agency. In this paper, we presented \textsc{Transit-Gym}, a SUMO based transit simulator for simulating a variety of transit scenarios. Its domain-specific modeling language and associated experimentation tool-chain and infrastructure enables intuitive and rapid specification of scenario variations in transit demand, route networks, and vehicle trip assignments as well as simulating them on integrated cloud backend. We also presented a novel approach to perform stochastic checks and calibrate crowdsourced street maps and traffic demand patterns for accurate testing and validation. Our experiment results demonstrated the broader applicability of \textsc{Transit-Gym} for realistic transit simulations and providing efficient decision-support for transportation planning. \textsc{Transit-Gym} is freely available on GitHub: \textit{https://github.com/smarttransit-ai/transit-simulator}. 

We are currently working on several extensions to \textsc{Transit-Gym} including: (1) integrating transportation planning with transit simulation and energy consumption estimation together as a co-simulation using Vanderbilt's CPSWT framework, (2) developing a richer scenario modeling language to incorporate more specific transit use-cases, (3) increasing performance of our simulation through partitioned traffic simulations, (4) scaling our simulations for broader cloud integration, (5) incorporating certain sensitive attributes at the passengers to evaluate whether or not the simulation scenarios lead to equitable  and  fair  coverage  to  areas  with  low  rider-ship, and (6) enhancing visualization of experiment and analysis results.
\section{Acknowledgements}
\label{sec:acknowledgements}

This material is based upon work supported by National Science Foundation under grants CNS-1952011, CNS-2029950 and Department of Energy, Office of Energy Efficiency and Renewable Energy (EERE), under Award Number DEEE0008467. Any opinions, findings, and conclusions or recommendations expressed in this material are those of the author(s) and do not necessarily reflect the views of the National Science Foundation or the Department of Energy.

\bibliographystyle{IEEEtran}
\bibliography{main}

\begin{thebibliography}{10}
\providecommand{\url}[1]{#1}
\csname url@samestyle\endcsname
\providecommand{\newblock}{\relax}
\providecommand{\bibinfo}[2]{#2}
\providecommand{\BIBentrySTDinterwordspacing}{\spaceskip=0pt\relax}
\providecommand{\BIBentryALTinterwordstretchfactor}{4}
\providecommand{\BIBentryALTinterwordspacing}{\spaceskip=\fontdimen2\font plus
\BIBentryALTinterwordstretchfactor\fontdimen3\font minus
  \fontdimen4\font\relax}
\providecommand{\BIBforeignlanguage}[2]{{%
\expandafter\ifx\csname l@#1\endcsname\relax
\typeout{** WARNING: IEEEtran.bst: No hyphenation pattern has been}%
\typeout{** loaded for the language `#1'. Using the pattern for}%
\typeout{** the default language instead.}%
\else
\language=\csname l@#1\endcsname
\fi
#2}}
\providecommand{\BIBdecl}{\relax}
\BIBdecl

\bibitem{phillips2004application}
J.~K. Phillips, ``An application of the balanced scorecard to public transit
  system performance assessment,'' \emph{Transportation Journal}, pp. 26--55,
  2004.

\bibitem{murray2003coverage}
A.~T. Murray, ``A coverage model for improving public transit system
  accessibility and expanding access,'' \emph{Annals of Operations Research},
  vol. 123, no.~1, pp. 143--156, 2003.

\bibitem{lai2011behavioral}
W.-T. Lai and C.-F. Chen, ``Behavioral intentions of public transit
  passengers—the roles of service quality, perceived value, satisfaction and
  involvement,'' \emph{Transport policy}, vol.~18, no.~2, pp. 318--325, 2011.

\bibitem{sayarshad2020optimizing}
H.~R. Sayarshad and H.~O. Gao, ``Optimizing dynamic switching between fixed and
  flexible transit services with an idle-vehicle relocation strategy and
  reductions in emissions,'' \emph{Transportation Research Part A: Policy and
  Practice}, vol. 135, pp. 198--214, 2020.

\bibitem{sivagnanam2021minimizing}
A.~Sivagnanam, A.~Ayman, M.~Wilbur, P.~Pugliese, A.~Dubey, and A.~Laszka,
  ``Minimizing energy use of mixed-fleet public transit for fixed-route
  service,'' in \emph{35th AAAI Conference on Artificial Intelligence
  (AAAI-21)}, February 2021.

\bibitem{gallet2018estimation}
M.~Gallet, T.~Massier, and T.~Hamacher, ``Estimation of the energy demand of
  electric buses based on real-world data for large-scale public transport
  networks,'' \emph{Applied energy}, vol. 230, pp. 344--356, 2018.

\bibitem{ayman_data}
A.~Ayman, A.~Sivagnanam, M.~Wilbur, P.~Pugliese, A.~Dubey, and A.~Laszka,
  ``Data-driven prediction and optimization of energy use for transit fleets of
  electric and {ICE} vehicles,'' \emph{ACM Transactions on Internet Technology,
  in press}, vol. in press.

\bibitem{Samal2019}
C.~Samal, A.~Dubey, and L.~J. Ratliff, ``Mobilytics-{Gym}: {A} simulation
  framework for analyzing urban mobility decision strategies,'' in \emph{5th
  {IEEE} International Conference on Smart Computing}, June 2019, pp. 283--291.

\bibitem{haklay2010good}
M.~Haklay, ``How good is volunteered geographical information? a comparative
  study of openstreetmap and ordnance survey datasets,'' \emph{Environment and
  planning B: Planning and design}, vol.~37, no.~4, pp. 682--703, 2010.

\bibitem{sztipanovits1997model}
J.~Sztipanovits and G.~Karsai, ``{Model-Integrated Computing},''
  \emph{Computer}, vol.~30, no.~4, pp. 110--111, 1997.

\bibitem{kiczales1996aspect}
G.~Kiczales, ``Aspect-oriented programming,'' \emph{ACM Computing Surveys
  (CSUR)}, vol.~28, no. 4es, pp. 154--es, 1996.

\bibitem{batory1997composition}
D.~Batory and B.~J. Geraci, ``Composition validation and subjectivity in
  genvoca generators,'' \emph{IEEE Transactions on Software Engineering},
  vol.~23, no.~2, pp. 67--82, 1997.

\bibitem{mda}
{Object Management Group}, ``Model-driven architecture,''
  \url{http://www.omg.org/mda}.

\bibitem{neema2019simulation}
H.~Neema, J.~Sztipanovits, C.~Steinbrink, T.~Raub, B.~Cornelsen, and
  S.~Lehnhoff, ``Simulation integration platforms for cyber-physical systems,''
  in \emph{Workshop on Design Automation for CPS and IoT}, 2019, pp. 10--19.

\bibitem{koutsoukos2017sure}
X.~Koutsoukos, G.~Karsai, A.~Laszka, H.~Neema, B.~Potteiger, P.~Volgyesi,
  Y.~Vorobeychik, and J.~Sztipanovits, ``Sure: A modeling and simulation
  integration platform for evaluation of secure and resilient cyber--physical
  systems,'' \emph{Proceedings of the IEEE}, vol. 106, no.~1, pp. 93--112,
  2017.

\bibitem{neema2014model}
H.~Neema, J.~Gohl, Z.~Lattmann, J.~Sztipanovits, G.~Karsai, S.~Neema, T.~Bapty,
  J.~Batteh, H.~Tummescheit, and C.~Sureshkumar, ``Model-based integration
  platform for fmi co-simulation and heterogeneous simulations of
  cyber-physical systems,'' in \emph{Proceedings of the 10 th International
  Modelica Conference; March 10-12; 2014; Lund; Sweden}, no. 096.\hskip 1em
  plus 0.5em minus 0.4em\relax Link{\"o}ping University Electronic Press, 2014,
  pp. 235--245.

\bibitem{neema2016c2wt}
H.~Neema, J.~Sztipanovits, M.~Burns, and E.~Griffor, ``{C2WT-TE}: A model-based
  open platform for integrated simulations of transactive smart grids,'' in
  \emph{2016 Workshop on Modeling and Simulation of Cyber-Physical Energy
  Systems (MSCPES)}.\hskip 1em plus 0.5em minus 0.4em\relax IEEE, 2016, pp.
  1--6.

\bibitem{burns2018universal}
M.~Burns, T.~Roth, E.~Griffor, P.~Boynton, J.~Sztipanovits, and H.~Neema,
  ``{Universal CPS Environment for Federation (UCEF)},'' in \emph{2018 Winter
  Simulation Innovation Workshop}, 2018.

\bibitem{neema2018integrated}
H.~Neema, B.~Potteiger, X.~Koutsoukos, G.~Karsai, P.~Volgyesi, and
  J.~Sztipanovits, ``Integrated simulation testbed for security and resilience
  of cps,'' in \emph{Proceedings of the 33rd Annual ACM Symposium on Applied
  Computing}, 2018, pp. 368--374.

\bibitem{krajzewicz2012recent}
D.~Krajzewicz, J.~Erdmann, M.~Behrisch, and L.~Bieker, ``Recent development and
  applications of sumo-simulation of urban mobility,'' \emph{International
  journal on advances in systems and measurements}, vol.~5, no. 3\&4, 2012.

\bibitem{aziz2020optimization}
H.~A. Aziz, V.~Garikapati, T.~K. Rodriguez, L.~Zhu, B.~Sun, S.~E. Young, and
  Y.~Chen, ``An optimization-based planning tool for on-demand mobility service
  operations,'' \emph{International Journal of Sustainable Transportation}, pp.
  1--12, 2020.

\bibitem{becker2020assessing}
H.~Becker, M.~Balac, F.~Ciari, and K.~W. Axhausen, ``Assessing the welfare
  impacts of shared mobility and mobility as a service (maas),''
  \emph{Transportation Research Part A: Policy and Practice}, vol. 131, pp.
  228--243, 2020.

\bibitem{zhu2021shared}
L.~Zhu, Z.~Zhao, and G.~Wu, ``Shared automated mobility with demand-side
  cooperation: A proof-of-concept microsimulation study,''
  \emph{Sustainability}, vol.~13, no.~5, p. 2483, 2021.

\bibitem{masikos2015mesoscopic}
M.~Masikos, K.~Demestichas, E.~Adamopoulou, and M.~Theologou, ``Mesoscopic
  forecasting of vehicular consumption using neural networks,'' \emph{Soft
  Computing}, vol.~19, no.~1, pp. 145--156, 2015.

\bibitem{pamula2020estimation}
T.~Pamu{\l}a and W.~Pamu{\l}a, ``Estimation of the energy consumption of
  battery electric buses for public transport networks using real-world data
  and deep learning,'' \emph{Energies}, vol.~13, no.~9, p. 2340, 2020.

\bibitem{sun2019machine}
S.~Sun, J.~Zhang, J.~Bi, and Y.~Wang, ``A machine learning method for
  predicting driving range of battery electric vehicles,'' \emph{Journal of
  Advanced Transportation}, vol. 2019, 2019.

\bibitem{genikomsakis2017computationally}
K.~N. Genikomsakis and G.~Mitrentsis, ``A computationally efficient simulation
  model for estimating energy consumption of electric vehicles in the context
  of route planning applications,'' \emph{Transportation Research Part D:
  Transport and Environment}, vol.~50, pp. 98--118, 2017.

\bibitem{luin2019microsimulation}
B.~Luin, S.~Petelin, and F.~Al-Mansour, ``Microsimulation of electric vehicle
  energy consumption,'' \emph{Energy}, vol. 174, pp. 24--32, 2019.

\bibitem{sun2021hybrid}
R.~Sun, Y.~Chen, A.~Dubey, and P.~Pugliese, ``Hybrid electric buses fuel
  consumption prediction based on real-world driving data,''
  \emph{Transportation Research Part D: Transport and Environment}, vol.~91, p.
  102637, 2021.

\bibitem{wegener2008traci}
A.~Wegener, M.~Pi{\'o}rkowski, M.~Raya, H.~Hellbr{\"u}ck, S.~Fischer, and J.-P.
  Hubaux, ``Traci: an interface for coupling road traffic and network
  simulators,'' in \emph{Proceedings of the 11th communications and networking
  simulation symposium}, 2008, pp. 155--163.

\end{thebibliography}

\end{document}